\documentclass[showpacs,preprintnumbers,amssymb]{revtex4-2}
\usepackage{graphicx}
\usepackage{dcolumn}
\usepackage{color}
\usepackage{bm}
\usepackage{amsmath}
\usepackage{amssymb}
\usepackage{epsfig}
\usepackage{epsfig}
\usepackage{epstopdf}
\usepackage{multirow}

\def \a{\alpha}
\def \b{\beta}
\def \l{\lambda}

\def \d{\delta}
\def \k{\kappa}

\def \be{\begin{equation}}
\def \ee{\end{equation}}
\def \ben{\begin{eqnarray}}
\def \een{\end{eqnarray}}

\def \O{\Omega}

\def \S{\Sigma}
\def \p{\partial}
\def \t{\theta}
\def \P{\Phi}
\def \M{\mathcal{M}}
\def \G{\mathcal{G}}
\def \r{\rho}
\def \E{\mathcal{E}}
\def \K{\mathcal{K}}
\def \R{\mathcal{R}}
\def \m{\mathbb{M}}
\def \U{\mathcal{U}}
\def \L{\mathcal{L}}
\def \F{\mathcal{F}}
\def \gg{\mathbb{G}}

\def\be{\begin{equation}}
\def\ee{\end{equation}}
\def\bse{\begin{subequations}}
\def\ese{\end{subequations}}
\def\ba{\begin{array}}
\def\ea{\end{array}}
\def\bea{\begin{eqnarray}}
\def\eea{\end{eqnarray}}

\begin{document}
	
\title{Collapsing scenario for the {\bf k}-essence emergent generalised Vaidya spacetime in the context of massive gravity's rainbow}

\author{Saibal Ray }
\altaffiliation{saibal.ray@gla.ac.in}
\affiliation{Centre for Cosmology, Astrophysics and Space Science (CCASS), GLA University, Mathura 281406, Uttar Pradesh, India}
  
\author{Arijit Panda$^a$}
\altaffiliation{arijitpanda260195@gmail.com}
\affiliation{Department of Physics, Prabhat Kumar College, Contai, Purba Medinipur-721404, India}

\author{Bivash Majumder}
\altaffiliation{bivashmajumder@gmail.com}
\affiliation{Department of Mathematics, Prabhat Kumar College, Contai, Purba Medinipur-721404, India}

\author{Md. Rabiul Islam}
\altaffiliation{rabi76.mri@gmail.com}
\affiliation{$^a$ Department of Physics, Raiganj University, Raiganj, Uttar Dinajpur-733 134, West Bengal, India.}

\author{Goutam Manna$^b$}
\altaffiliation{goutammanna.pkc@gmail.com\\ $^b$ Corresponding author}
\affiliation{Department of Physics, Prabhat Kumar College, Contai, Purba Medinipur-721404, India}

\begin{abstract}
In this paper, we study the collapsing scenario for the {\bf k-}essence emergent Vaidya spacetime in the context of massive gravity's rainbow. For this study, we consider that the background metric is Vaidya spacetime in massive gravity's rainbow. We show that the {\bf k-}essence emergent gravity metric resembles closely to the new type of generalized Vaidya massive gravity metric with the rainbow deformations for null fluid collapse where we consider the {\bf k-}essence scalar field as a function solely of the advanced or the retarded time. The {\bf k-}essence emergent Vaidya massive gravity rainbow mass function is also different. This new type {\bf k-}essence emergent Vaidya massive gravity rainbow metric has satisfied the required energy conditions. The existence of the locally naked central singularity, the strength and the strongness of the singularities for the rainbow deformations of the {\bf k-}essence emergent Vaidya massive gravity metric are the interesting outcomes of the present work.
\end{abstract}



\maketitle

\section{Introduction}\label{intro}
Based on the nonlinear or doubly special relativity (DSR)~\cite{magueijo2,magueijo3,magueijo4}, Magueijo et al.~\cite{magueijo1} were the first to develop the rainbow gravity. Kimberly et al.~\cite{kim} discussed the details of this gravity, which is basically a deformed version of the special relativity (SR). One of the major characteristics of this theory is that, it was proposed to keep inertial frames relative. Secondly, Planck energy can be made of an invariant scale~\cite{magueijo1,cam,ling} by this theory. Here comes the proposal of a dual to non-linear realization of relativity in the momentum space resulting, the spacetime invariant, an energy-dependent metric. It is noteworthy that, there exist a variety of theories in literature where, the modification of  standard energy-momentum dispersion relation has been done in the limit of the Planck scale, such as, string theory~\cite{kost}, loop quantum gravity~\cite{gam} and non-commutative geometry~\cite{carr}. However, to assimilate the curvature of the rainbow gravity~\cite{magueijo1} generalizes the DSR taking the energy-dependency of spacetime into consideration. This type of consideration permits the quanta of different energies to observe different classical geometries. The argument for the allocation of same inertial frames is that, we want to conserve the equivalence principle in a modified frame and that's because all the measurements of distance and time become dependent on the energy of the quanta, which are to be engaged for the testing purposes.

The deformed dispersion relations, in the Planck length scale are, $E^{2}f^{2}(E)-(\overrightarrow{p})^{2}g^{2}(E)=m^{2}$~\cite{magueijo1}. However, the standard linear Lorentz transformations do not ensure the invariance of these relations. Whether the invariance of these have been proved under non-linear Lorentz transformations by Kimberly et al.~\cite{kim}. It has already been acknowledged that, in place of single spacetime, dual to momentum space is actually the energy-dependent family of the metric. A notable point is that $E$ in the metric $g_{\mu\nu} (E)$ has nothing to do with the energy of the spacetime, instead it is the scale at which the geometry of spacetime is probed. Magueijo et. al.~\cite{magueijo1} argued that seeing a particle (or plane wave or wave-packet) with energy $E$ by an observer means that particular particle experiences the metric $g_{\mu\nu}(E)$. Moreover, if this particle is seen with different $E'(\neq E)$ by a different observer, then a different metric will be assigned to the particle, i.e. $g_{\mu\nu}(E')$. The requirement of the covariance for a non-linear representation of the Lorentz group in momentum space may demand, seeing a given particle being affected by different metrics by different observers, as well as, assigning different metrics to different particles moving in the same region at the same time by same observer.  

The modification of the equivalence principle have also been established by Magueijo et al.~\cite{magueijo1}, which states that, the spacetime is described by a one-parameter family of metrics given in terms of a one-parameter family of orthonormal frame fields as $g(E)=\eta^{\mu\nu}e_{\mu}(E)\otimes e_{\nu}(E)$, where the frame fields depends on the energy as, $e_{0}(E)=\frac{1}{f(E)}\tilde{e}_{0}$ and $e_{i}(E)=\frac{1}{g(E)}\tilde{e}_{i}$ in the Planck length scale. Here the tilde bearing terms refer to the energy-independent frame fields, that specify the geometry probed by the low energy quanta, the limit being $El_{P}<<1$ with $g_{\mu\nu}\simeq \eta_{\mu\nu}$, where $l_{P}$ is the Planck length. 

However, an interesting point, which we would like to mention here that, if the energy $E$ is just a non-dynamical parameter of the theory then, rescaling can gauge it away easily. Unfortunately, the dependence on the coordinate is dynamical and actually breaks the diffeomorphisms symmetry of the whole metric. By the fact that, the local symmetry in gravity’s rainbow is not Lorentz symmetry, because it depends on the deformation of the usual energy-momentum dispersion relation and consequently breaks the Lorentz symmetry in the UV limit also~\cite{hendi6}. We should also remember that, the energy $E$ is an implicit function of the coordinates whereas, rainbow functions are explicit functions of the energy $\E$ which is defined as the ratio $E/E_P$, $E_P$ being the Planck energy. That immediately concludes that, these are also dynamical functions of the coordinates, restricting us to gauge these away. The difficulty to find such an explicit dependence of rainbow functions on energy for different systems also goes in vain because, these are implicit dynamical functions of the coordinates, that means, gauging away is not going to entertain us anyway. So we can expect that these functions may produce physically different results from general relativity~\cite{ali5}.

Ali et al.~\cite{ali1} have shown that the gravitational collapse can be explained in the context of gravity's rainbow. Heydarzade et. al.~in \cite{heyd1,heyd2} have discussed the energy dependent deformation and time dependent geometry of massive gravity with the help of massive gravity's rainbow formalism. Vainshtein and the dRGT mechanisms have been used side by side for the energy dependent massive gravity and additionally their works includes the analysis of a ghost free theory of massive gravity's rainbow with the help of the radiating Vaidya solution. Phenomenologically, the massive gravity can also be motivated from accelerated cosmic expansion~\cite{riess1,riess2,riess3,perl1,perl2,tonry}. The problems with the massive gravity can be resolved using the Vainshtein mechanism~\cite{vain,babichev} easily. However, the Vainshtein mechanism can raise the Boulware-Deser ghosts~\cite{boul}, which again can be resolved by using the dRGT mechanism~\cite{rham1,rham2,rham3,hassan1,hassan2,hassan3,hassan4,hint}.

Some related theories of rainbow gravity and their subsequent investigation in charged dilatonic black holes, Gauss-Bonnet gravity, Lovelock gravity, combination of Rastall and rainbow theories, $AdS_{4}$ dyonic black holes, deformed Starobinsky model, thermodynamics of black holes, Galileon gravity, horizon effect, violation of weak cosmic censorship and Branes are addressed in~\cite{ali2,ali3,ali4,hendi1,hendi2,hendi3,hendi4,hendi5,rudra,ashour,hendi6,mota,pana,chan,gim,haldar}.

Either a black hole or a naked singularity may be produced from Gravitational collapse~\cite{joshi1,joshi2}. Also, it may provide mechanism to communicate with far-away observers in the universe. The radiating Schwarzschild spacetime also known as the Vaidya spacetime~\cite{vaidya1} can describe the geometry outside a radiating spherically symmetric star. Particularly, the formation of naked singularities can be produced from the solution of the null dust fluid with spherical symmetry in gravitational collapse, as shown by Papapetrou~\cite{papa}. Additionally, cosmic censorship conjecture (CCC)~\cite{penrose} can be encountered in a different way by this example. The causal trajectories joining the singularities in the ingoing Vaidya situation have been studied by the authors of~\cite{joshi3,vert}. They also provide the whole story of constraints and classification the non-spacelike geodesics that, make connection with the past naked singularity. Then, this naked singularity was shown to be a strong curvature singularity in a more effective sense. 

The solution of Vaidya spacetime~\cite{vai1,vai2,vai3,vai4,vai5,jg} was generalised including all the possible solutions of Einstein's field equations and combining Type-I and Type-II matter fields in the works of Husain~\cite{husain} and Wang and Wu~\cite{wang}. the generalization of the Vaidya solution in the context of the cosmic censorship hypothesis, which is basically the study of the gravitational collapse for the generalized Vaidya spacetime has been done in~\cite{maombi}. It was their findings that, to terminate the collapse the classes of generalized Vaidya mass functions should brought in the scenario with a locally naked central singularity. Along with they also calculated the strength of these singularities. For the termination of the non-spacelike geodesics directed towards the future at the singularity in the past they studied the conditions on the mass function, developing a general mathematical framework. The fact that, the final  outcome of collapse can be determined in terms of either a black hole or a naked singularity, for a given generalized Vaidya mass function was also established. The gravitational collapse of higher dimension, in the charged-Vaidya spacetime was studied by the authors in~\cite{patil}. Their works include that, singularities arising in a charged null fluid in a higher dimensional spacetime are always naked, violating strong cosmic censorship hypothesis (CCH). The holographic complexity and the holographic entropy cone in AdS-Vaidya spacetime was explored in~\cite{cag,momeni}. Sharma et al.~\cite{sharma} revisited the Vaidya-Tikekar stellar model in the linear regime. Manna et al. established  the general connection between the {\bf k}-essence geometry and the Vaidya spacetime in~\cite{gm4,gm6,gm5}. More specifically, using a generalized Vaidya-type metric~\cite{husain,wang}, they have worked on the gravitational collapse in the {\bf k}-essence emergent gravity  in~\cite{gm5} in the context of cosmic censorship hypothesis. Alongside, the existence of the locally naked central singularity, the strength, and the strongness of the singularities for the {\bf k}-essence emergent Vaidya metric was established by them.

The {\bf k-}essence model, particularly the scalar field model reveals that the role of kinetic energy is much more dominating than the potential energy of the field. The kinetic energy term does not depend explicitly on the field and therefore can not be separated in the theoretical form of Lagrangian with non-canonical kinetic terms theoretical form. Born and Infeld~\cite{born1,born2,born3} first proposed a theory with a non-canonical kinetic term which was related to the infinite self-energy of the electron.

The general form of the Lagrangian for {\bf k-}essence model can be written as: $L=-V(\phi)F(X)$ where $\phi$ is {\bf k}-essence scalar field, $X=\frac{1}{2}g^{\mu\nu}\nabla_{\mu}\phi\nabla_{\nu}\phi$ and shows the non-dependency on $\phi$ to start with~\cite{babi1,babi2,babi3,babi4,babi5,gm1,gm2,gm3,scherrer1,scherrer2,Tian}. The {\bf k-}essence theory with non-canonical kinetic terms and the relativistic field theories with canonical kinetic terms differs in the non-trivial dynamical solutions of the {\bf k}-essence equation of motion (EOM). The metric for the perturbations around these solutions are changed by solutions of the EOM breaking the Lorentz invariance spontaneously. This allows the perturbations to propagate in the so-called emergent or analogue curved spacetime~\cite{babi1,babi2,babi3,babi4,babi5} with the metric, different from the gravitational one. The emergent gravity metric $\tilde G_{\mu\nu}$ is not conformally equivalent to the gravitational metric  $g_{\mu\nu}$.

Based on the following articles~\cite{heyd1,heyd2,gm5}, our work in this paper includes the exploration of the collapsing scenario for the {\bf k}-essence emergent Vaidya spacetime in the context of massive gravity's rainbow. To do this, we started our work by considering the background metric as Vaidya spacetime in massive gravity's rainbow~\cite{heyd1}. A notable thing to mention is that, the quantization problem of the gravity is beyond our work for this article. Our consideration for this work is that, the background metric is rainbow deformations of the massive gravity Vaidya type. Doing this, we have discussed the collapsing scenario on the basis of the {\bf k-}essence geometry. 

Let us now discuss some importance of the {\bf k}-essence theory as follows: The Coincidence problem has questioned most of the dark energy models (e.g., cosmological constant) in recent years by bringing some of the observational data (Large-Scale Structure, searches of type Ia Supernovae and measurements of Cosmic Microwave Background anisotropy~\cite{Bahcall}) in the scenario. The problem is the fine tuning of the initial energy density which is of the orders of 100 or more smaller than the initial matter-energy density. The solution of the abovementioned problem lies in the {\bf k}-essence theory which is a nonlinear kinetic energy of the scalar field~\cite{Picon1}. An elaborated discussion about the motivation of {\bf K}-essence theory has been given in Ref.~\cite{gm8}. Not only in cosmology~\cite{gm7,Nojiri,Pasqua,Socorro,Chimento} but also other fields of gravity has dealt with {\bf K}-essence theory~\cite{gm5,Akhoury,Gannouji}. All these aspects have motivated us to consider the issues in the present work. 

We embodied this paper as follows: The {\bf k}-essence geometry based on the Dirac-Born-Infeld action, rainbow theory of gravity and massive gravity have been revisited in Sect. 2. In Sect. 3, we concentrate to build the rainbow deformations of the {\bf k}-essence emergent Vaidya massive gravity spacetime considering the background metric as Vaidya massive gravity's rainbow. A scalar field with a restriction of being an arbitrary function of the advanced or retarded Eddington-Finkelstein time has been considered where, the scalar field is independent of the other variables of the four-dimensional spacetime. The Ricci and Einstein tensors corresponding to our rainbow deformations of the {\bf k}-essence emergent Vaidya massive gravity metric has also been constructed. This section also include the computation of the components of the emergent energy-momentum tensor by direct substitution into an emergent Einstein equation. This emergent tensor must obey the energy conditions of the emergent geometry. In Sect. 4, we have developed the collapsing scenario for the {\bf k-}essence emergent Vaidya spacetime in massive gravity's rainbow. In this section, we have analyzed the structure of the central singularity to find out the conditions on the {\bf k-}essence emergent Vaidya massive gravity's rainbow mass function with the rainbow functions, existence of outgoing nonspacelike geodesic and the strength of the singularities for the above {\bf k-}essence emergent geometry. The last Sect. 5 contains some conclusion as well as discussion of our work.

\section{Brief review of the K-essence geometry, gravity's rainbow theory and massive gravity}

\subsection{The K-essence geometry}

The action of the {\bf k}-essence scalar field $\phi$, minimally coupled to the background spacetime metric $g_{\mu\nu}$ is given by~\cite{babi1}--\cite{babi5}
\be
S_{k}[\phi,g_{\mu\nu}]= \int d^{4}x {\sqrt -g} L(X,\phi), \label{1}
\ee
where $X={1\over 2}g^{\mu\nu}\nabla_{\mu}\phi\nabla_{\nu}\phi$.

The energy-momentum tensor can be written as
\be
T_{\mu\nu}\equiv {2\over \sqrt {-g}}{\delta S_{k}\over \delta g^{\mu\nu}}= L_{X}\nabla_{\mu}\phi\nabla_{\nu}\phi - g_{\mu\nu}L, \label{2}
\ee
where $L_{\mathrm X}= {dL\over dX},~L_{\mathrm XX}= {d^{2}L\over dX^{2}},~L_{\mathrm\phi}={dL\over d\phi}$ and  $\nabla_{\mu}$ is the covariant derivative defined with respect to the gravitational metric $g_{\mu\nu}$.

The equation of motion of the scalar field is
\be
-{1\over \sqrt {-g}}{\delta S_{k}\over \delta \phi}= \tilde G^{\mu\nu}\nabla_{\mu}\nabla_{\nu}\phi +2XL_{X\phi}-L_{\phi}=0, \label{3}
\ee
where
\be
\tilde G^{\mu\nu}\equiv L_{X} g^{\mu\nu} + L_{XX} \nabla ^{\mu}\phi\nabla^{\nu}\phi \label{4}
\ee
and $1+ {2X  L_{XX}\over L_{X}} > 0$.

The conformal transformation gives
$G^{\mu\nu}\equiv {c_{s}\over L_{x}^{2}}\tilde G^{\mu\nu}$, with $c_s^{2}(X,\phi)\equiv{(1+2X{L_{XX}\over L_{X}})^{-1}}$, the inverse metric of $G^{\mu\nu}$ takes the form
\be
G_{\mu\nu}={L_{X}\over c_{s}}[g_{\mu\nu}-{c_{s}^{2}}{L_{XX}\over L_{X}}\nabla_{\mu}\phi\nabla_{\nu}\phi]. \label{5}
\ee

Again, applying a conformal transformation~\cite{gm1,gm2} $\bar G_{\mu\nu}\equiv {c_{s}\over L_{X}}G_{\mu\nu}$, we get
\be 
\bar {G}_{\mu\nu}
={g_{\mu\nu}-{{L_{XX}}\over {L_{X}+2XL_{XX}}}\nabla_{\mu}\phi\nabla_{\nu}\phi}.\label{6}
\ee	

To make Eqs. (1)--(4) physically meaningful, we should have $L_{X}\neq 0$ for $c_{s}^{2}$ to be positive definite. If $L$ does not directly depend on $\phi$, then the EOM (3) reduces to
\be
-{1\over \sqrt {-g}}{\delta S_{k}\over \delta \phi}
= \bar {G}^{\mu\nu}\nabla_{\mu}\nabla_{\nu}\phi=0.\label{7}
\ee

The Dirac-Born-Infeld (DBI) type Lagrangian~\cite{born1,born2,born3,gm1,gm2,gm3} has the form
\be
L(X,\phi)= 1-V(\phi)\sqrt{1-2X},
\label{8}
\ee
with $V(\phi)=V=$constant and kinetic energy of $\phi>>V$, i.e .$(\dot\phi)^{2}>>V$. This ensures the domination of kinetic energy dominates over the potential energy for the {\bf k}-essence fields and gives us $c_{s}^{2}(X,\phi)=1-2X$. For scalar fields $\nabla_{\mu}\phi=\partial_{\mu}\phi$. Therefore, the effective emergent metric (\ref{6}) ends up as
\be
\bar{G}_{\mu\nu}= g_{\mu\nu} - \partial _{\mu}\phi\partial_{\nu}\phi.
\label{9}
\ee

The new Christoffel symbols and old Christoffel symbols are related to each other~\cite{wald,gm1} as

\be
\bar\Gamma ^{\alpha}_{\mu\nu} 
=\Gamma ^{\alpha}_{\mu\nu} -\frac {1}{2(1-2X)}[\delta^{\alpha}_{\mu}\partial_{\nu}X
+ \delta^{\alpha}_{\nu}\partial_{\mu}X].\label{10}
\ee
 
Now, we can write the new geodesic equation for the {\bf k-}essence theory in terms of the new Christoffel connections  $\bar\Gamma$ as
\be
\frac {d^{2}x^{\alpha}}{d\lambda^{2}} +  \bar\Gamma ^{\alpha}_{\mu\nu}\frac {dx^{\mu}}{d\lambda}\frac {dx^{\nu}}{d\lambda}=0, \label{11}
\ee
where $\l$ is an affine parameter.

\subsection{Gravity's rainbow theory}
Inspired by the works of Magueijo et al.~\cite{magueijo1,magueijo2,magueijo3,magueijo4} as well as Kimberly et al.~\cite{kim}, the deformed energy-momentum dispersion relation can be provided as
\be
E^{2}\F^{2}(\E)-p^{2}\G^{2}(\E)=m^{2}, \label{12}
\ee
where $\E=\frac{E}{E_{P}}$ is the energy ratio, which is obviously a dimensionless quantity, $E$ and $p$ has been used to express the energy and momentum (respectively) of the test particle and $E_{P}$ as the Planck energy. 

The fact that, the energy of a test particle cannot exceed the Plank energy, gives us the limit, $\E$ as $0<\E\leq 1$. Therefore, the  energy dependent rainbow functions $\mathcal{F}(\E)$ and $\mathcal{G}(\E)$ will satisfy the below two conditions:
\be
\lim_{\mathcal{E}\rightarrow 0}\mathcal{F}(\mathcal{E})=1~\text{and}~\lim_{\mathcal{E}\rightarrow 0}\mathcal{G}(\mathcal{E})=1 \label{13}
\ee
and the general relativity is recovered in the IR limit of the theory~\cite{Galan,Hackett,Girelli,Liu,Ling,Garattini,Peng}.

Again, the energy dependent contribution in the metric is given by
\be
g(\E)=\eta^{\mu\nu}e_{\mu}(\E)\otimes e_{\nu}(\E),\label{14}
\ee
with the energy dependence of the frame fields as follows: $e_{0}(\E)=\frac{1}{f(\E)}\tilde{e}_{0}$ and 
$e_{i}(\E)=\frac{1}{g(\E)}\tilde{e}_{i}$ in the Planck length scale. Here  $\tilde{e}_{0}$  and $\tilde{e}_{i}$ are the energy independent frame fields.

The choice of the rainbow functions can be of various type~\cite{camel1,camel2,heyd1,ali1} such as,  from the background motivation of loop quantum gravity and $\kappa-$Minkowski noncommutative spacetime:
\be
\F(\E)=1~\text{and}~~~\G(\E)=\sqrt{1-\a\E^{q}},\label{15}
\ee
where we can take the rainbow functions with the constant velocity of light \cite{magueijo4} as
\be
\F(\E)=\G(\E)=\frac{1}{1-\a\E}. \label{16}
\ee

For the hard spectra from gamma-ray burster’s at cosmological distances, it is also possible to choose rainbow functions~\cite{camel3} as
\be
\F(\E)=\frac{e^{\a\E}-1}{\a\E}~\text{and}~\G(\E)=1.\label{17}
\ee

Whatever be the choice, the main property of all these rainbow functions are spacetime energy dependent.

\subsection{Massive gravity}
The 4-dimensional massive gravity action~\cite{heyd1,heyd2,hendi7,hendi8} can be written as
\be
S=\int d^{4}x\sqrt{-g}\Big[ \R+ \m^{2}\sum_{i}^{4}c_{i}\mathcal{U}_{i}(g,f)+\L_{m}\Big],\label{18}
\ee
$f$ being a fixed symmetric tensor (also known as the reference metric), $c_{i}$ are constants, $\m$ are the massive gravity parameter and $\mathcal{U}_{i}$ are symmetric polynomials of the eigenvalues of the $d\times d$ matrix $\K^{\mu}_{\nu}=\sqrt{g^{\mu \alpha}f_{\alpha \nu}}$. 

The symmetric polynomials, mentioned above, can be written as
\bea
\mathcal{U}_{1}&=&[\K],\nonumber\\
\mathcal{U}_{2}&=&[\K]^{2}-[\K^{2}],\nonumber\\
\mathcal{U}_{3}&=&[\K]^{3}-3[\K][\K^{2}]+2[\K^{3}],\nonumber\\
\mathcal{U}_{4}&=&[\K]^{4}-6[\K^{2}][\K]^{2}+8[\K^{3}][\K]+3[\K^{2}]^{2}-6[\K^{4}],\nonumber\\\label{19}
\eea
where $\mathcal{K}=\mathcal{K}^{\mu}_{\mu}$. 

With the help of the variational principle the modified field equations for the massive gravity can be obtained as 
\be
\R_{\mu\nu}-\frac{1}{2}\R g_{\mu\nu}+\m^{2}\chi_{\mu\nu}=T_{\mu\nu}.\label{20}
\ee

Here, $\chi_{\mu\nu}$ denotes the massive term, which can be expressed as
\ben
\chi_{\mu\nu}&=&-\frac{c_{1}}{2}\Big(\U_{1}g_{\mu\nu}-\K_{\mu\nu}\Big)-\frac{c_{2}}{2}\Big(\U_{2}g_{\mu\nu}-2\U_{1}\K_{\mu\nu}+2\K^{2}_{\mu\nu}\Big)\nonumber\\&-&\frac{c_{3}}{2}\Big(\U_{3}g_{\mu\nu}-3\U_{2}\K_{\mu\nu}+6\U_{1}\K^{2}_{\mu\nu}-6\K^{3}_{\mu\nu}\Big)\nonumber\\&-&\frac{c_{4}}{2}\Big(\U_{4}g_{\mu\nu}-4\U_{3}\K_{\mu\nu}+12\U_{2}\K^{2}_{\mu\nu}-24\U_{1}\K^{3}_{\mu\nu}+24\K^{4}_{\mu\nu}\Big), \label{21}
\een
considering $8\pi G=1$ in the geometrized units. 

Heydarzade et al.~\cite{heyd2} in their paper took a spatial reference metric in the basis of $(v,r,\theta,\Phi)$~\cite{heyd1,cai,vegh}
\be
f_{\mu\nu}=diag(0,0,c^{2}h_{ij}),\label{22}
\ee
in which $c^{2}$ is a positive constant and $h_{ij}$ is the two dimensional Euclidean metric.

We can write the barotropic relation and the energy-momentum tensor respectively as
\be
p=\kappa \rho,\label{23}
\ee
\be
T_{\mu\nu}=T^{(n)}_{\mu\nu}+T^{(m)}_{\mu\nu},\label{24}
\ee
with
\ben
T^{(n)}_{\mu\nu}&=&\sigma l_{\mu}l_{\nu},\nonumber\\
T^{(m)}_{\mu\nu}&=&(\rho+p)(l_{\mu}n_{\nu}+l_{\nu}n_{\mu})+p~{g}_{\mu\nu}.\label{25}
\een

The terms $T^{(n)}_{\mu\nu}$ and $T^{(m)}_{\mu\nu}$,  in the above expression are the energy-momentum tensor for the Vaidya null radiation and the energy-momentum tensor of the perfect fluid, respectively. $\sigma$, $\rho$ and $p$ denoting the null radiation density, energy density and pressure of the perfect fluid, respectively, and $l_{\mu}$, $n_{\mu}$ are the two null vectors. These null vectors are defined as \cite{heyd2}: $l_{\mu}=(1,0,0,0)$, $n_{\mu}=\Big(\frac{1}{2}(1-\frac{m(v,r)}{r}),-1,0,0\Big)$ with $l_{\mu}l^{\mu}=n_{\mu}n^{\mu}=0$ and $l_{\mu}n^{\mu}=-1$.

Taking the above assumptions into consideration, Heydarzade et al.~\cite{heyd2} have constructed the following metric for the Vaidya spacetime in massive gravity:

\be
ds^{2}=-\Big(1-\frac{m(v,r)}{r}\Big)dv^{2}+2dv~dr+r^{2}d\Omega^{2},\label{26}
\ee
with $d\Omega^{2}=d\theta^{2}+\sin^{2}\theta d\Phi^{2}$, and
\ben
m(v,r)=\frac{r^{1-2\kappa}}{1-2\kappa}f_{1}(v)+f_{2}(v)-\frac{1}{2}\mathbb{M}^{2}c_{1}cr^{2}-\mathbb{M}^{2}c_{2}c^{2}r,\nonumber\\ \label{27}
\een
under the constraint $\kappa \neq \frac{1}{2}$, where $f_{1}(v)$ and $f_{2}(v)$ are arbitrary functions of $v$ and can be expressed as
\ben
&&f_{1}(v)=\rho(v,r)r^{2(1+\kappa)};~~~\sigma(v,r)=\frac{r^{-(1+2\kappa)}}{1-2\kappa}\dot{f}_{1}(v)+\frac{1}{r^{2}}\dot{f}_{2}(v),\label{28}
\een
where a 'dot' represents the derivative with respect to $v$. Here, $v$ represents the null coordinate corresponding to the Eddington advanced time with $r$ decreasing towards the future along a ray related to $v=$ constant.

We can establish the following metric with the help of~\cite{heyd2}, considering the rainbow deformations of the Vaidya spacetime in massive gravity~\cite{heyd1}
\bea
ds^{2}=-\frac{1}{\F^{2}(\E)}\Big(1-\frac{m(v,r)}{r}\Big)dv^{2}+\frac{2}{\F(\E)\G(\E)}dvdr+\frac{1}{\G^{2}(\E)}r^{2}d\O^{2},\label{29}
\eea
with
\bea
m(v,r)&=&\frac{r^{1-2\k}}{1-2\k}f_{1}(v)+f_{2}(v)-\frac{1}{2\G(\E)}\m^{2}c_{1}cr^{2}-\m^{2}c_{2}c^{2}r,~~~(\kappa \neq\frac{1}{2}) \label{30}
\eea
and
\bea
f_{1}(v)=\frac{\rho(v,r)}{\G^{2}(\E)}r^{2(1+\k)};~~~~
\sigma(v,r)=\F(\E)\G(\E)\Big[\frac{r^{-(1+2\k)}}{1-2\k}\dot{f_{1}(v)}+\frac{1}{r^{2}}\dot{f_{2}(v)}\Big]. \label{31}
\eea

As expected, the mass function, given in (\ref{30}) is different from the mass function of Heydarzade et al.~\cite{heyd1}

\section{K-essence Vaidya spacetime in massive gravity's rainbow}
In this section, we will be busy to discuss the {\bf k-}essence Vaidya geometry in the context of the massive gravity's rainbow. Manna et al. have established the connection between the {\bf k}-essence geometry and the Vaidya spacetime based on the DBI type action in~\cite{gm4,gm5}.

So, let's assume the background metric to be the Vaidya massive gravity's rainbow (\ref{29}) with the definitions (\ref{30}) and (\ref{31}).

The {\bf k}-essence emergent line element can be written from Eq. (\ref{9})
\be
dS^{2}=ds^{2}-\p_{\mu}\phi\p_{\nu}\phi dx^{\mu}dx^{\nu}.\label{33}
\ee

Assuming the {\bf k}-essence scalar field $\phi(x)=\phi(v)$ only, the emergent line element in the context of massive gravity's rainbow (\ref{33}) can be written as
\bea
dS^{2}=-\Big[\frac{1}{\F^{2}(\E)}\Big(1-\frac{m(v,r)}{r}\Big)-\phi_{v}^{2}\Big]dv^{2}+\frac{2}{\F(\E)\G(\E)}dvdr+\frac{1}{\G^{2}(\E)}r^{2}d\O^{2},\label{34}
\eea
which also can be represented as~\cite{gm5}:
\bea
dS^{2}=-\frac{1}{\F^{2}(\E)}\Big(1-\frac{\M(v,r)}{r}\Big)dv^{2}+\frac{2}{\F(\E)\G(\E)}dvdr+\frac{1}{\G^{2}(\E)}r^{2}d\Omega^{2}.~~~~\label{35}
\eea

Here, we define the {\bf k}-essence emergent Vaidya massive gravity's rainbow mass function as
\bea
\mathcal{M}(v,r)&&=m(v,r)+r\F^{2}(\E)\phi_{v}^{2}\nonumber\\
&&=\frac{r^{1-2\k}}{1-2\k}f_{1}(v)+f_{2}(v)-\frac{1}{2\G(\E)}\m^{2}c_{1}cr^{2}-\m^{2}c_{2}c^{2}r+r\F^{2}(\E)\phi_{v}^{2},~~~~(\k\neq\frac{1}{2}),\label{36}
\eea
where $\phi_{v}^{2}$ ($\phi_{v}=\frac{\p\phi}{\p v}$) is the kinetic energy of the scalar field $\phi$, which {\it should not be equal to zero}.

For well-defined signature of the above metric (\ref{34}) the values of $\phi_{v}^{2}$ should lie between $0$ and $1$, i.e., $0<\phi_{v}^{2}<1$. Since, in general, spherical symmetry would only require that $\phi=\phi(v,r)$, the {\bf k-}essence scalar field $\phi$ actually violates local Lorentz invariance. The choice of $\phi(v,r)=\phi(v)$ additionally implies that outside of this particular choice of frame, a spherically symmetric $\phi$ is actually a function of both $v$ and $r$. Since, the dynamical solutions of the {\bf k-}essence equation of motion spontaneously break Lorentz invariance and also change the metric for the perturbations around these solutions, the {\bf k}-essence theory allows us to take this type of Lorentz violation. Also, notice that the choice of $\phi(v,r)~(=\phi(v))$ is equally important in the context of cosmology and gravitation \cite{gm4,gm6,gm5,gm7,gm8}.

Using the following definitions
\bea
\M_{v}=\dot{\M}(v,r)\equiv \frac{\p\M(v,r)}{\p v},~~~
\M_{r}=\M^{'}(v,r)\equiv\frac{\p\M(v,r)}{\p r},
\label{38}
\eea
the non-zero components of the emergent Ricci tensors can be represented as
\bea
\bar{R}^{v}_{v}=\bar{R}^{r}_{r}=\frac{\G^{2}(\E)}{2r}\M_{rr}=\frac{\G^{2}(\E)}{2r}m_{rr}~;~
\bar{R}^{\t}_{\t}=\bar{R}^{\P}_{\P}=\frac{\G^{2}(\E)}{r^{2}}\M_{r}=\frac{\G^{2}(\E)}{r^{2}}(m_{r}+\F^{2}(\E)\phi_{v}^{2}). \label{39}
\eea

Also the Ricci scalar is given by
\bea
\bar{R}=\frac{\G^{2}(\E)}{r}\M_{rr}
+\frac{2\G^{2}(\E)}{r^{2}}\M_{r}=\frac{\G^{2}(\E)}{r}m_{rr}
+\frac{2\G^{2}(\E)}{r^{2}}(m_{r}+\F^{2}(\E)\phi_{v}^{2}).
\label{40}
\eea

The non-vanishing components of the Einstein tensor are
\bea
\gg^{v}_{v}&=&\gg^{r}_{r}=-\frac{\G^{2}(\E)\M_{r}}{r^{2}}=-\frac{\G^{2}(\E)}{r^{2}}\Big(m_{r}+\F^{2}(\E)\phi_{v}^{2}\Big)~,~\nonumber\\
\gg^{r}_{v}&=&\frac{\G^{2}(\E)}{r^{2}}\M_{v}=\frac{\G^{2}(\E)}{r^{2}}\Big(m_{v}+2r\F^{2}(\E)\phi_{v}\phi_{vv}\Big)~,~\nonumber\\
\gg^{\t}_{\t}&=&\gg^{\P}_{\P}=-\frac{\G^{2}(\E)}{2r}\M_{rr}=-\frac{\G^{2}(\E)}{2r}m_{rr},
\label{41}
\een
where from Eq. (\ref{36}), we have
\ben
\M_{r}&=&f_{1}(v)r^{-2\k}-\frac{1}{\G(\E)}\m^{2}c_{1}cr-\m^{2}c_{2}c^{2}+\F^{2}(\E)\phi_{v}^{2}\label{42}\\
\M_{v}&=&\frac{r^{1-2\k}}{1-2\k}\dot{f_{1}(v)}+\dot{f_{2}(v)}+2r\F^{2}(\E)\phi_{v}\phi_{vv}\label{43}\\
\M_{rr}&=&-2\k r^{-(1+2\k)}f_{1}(v)-\frac{1}{\G(\E)}\m^{2}c_{1}c.\label{44}
\eea

The {\bf k}-essence emergent Einstein equation is
\be
\gg^{\mu}_{\nu} = {\cal T}^{\mu}_{\nu}.\label{45}
\ee

If we consider, $8\pi G=1$, then it immediately leads us to the components ${\cal T}^{\mu}_{ \nu}$, which can be parametrized exactly as in refs.~\cite{gm5,husain,wang,maombi} in terms of the components $\sigma, \rho$ and $p$ given by
\bea
{\cal T}_{\mu\nu}={\cal T}_{\mu\nu}^{(n)} + {\cal T}_{\mu\nu}^{(m)}=
\left[\begin{array}{cccc}
(\sigma/2+\r) & \sigma/2 & 0 & 0	\\
\sigma/2 & (\sigma/2-\r) & 0 & 0	\\
0 & 0 & p & 0	\\
0 & 0 & 0 & p
\end{array}\right], \label{46}
\eea
where ${\cal T}_{\mu\nu}^{(n)}=\sigma l_{\mu}l_{\nu}$;~ ${\cal T}_{\mu\nu}^{(m)}=(\r+p)(l_{\mu}n_{\nu}+l_{\nu}n_{\mu})+p\bar{G}_{\mu\nu}$ with $l_{\mu}$ and $n_{\mu}$ are two null vectors. The doubt in using the perfect fluid energy-momentum tensor for the {\bf k-}essence theory can be erased by the form of the Lagrangian $L(X) = 1-V\sqrt{1-2X}$, where $V$ is a constant, and it does not depend explicitly on $\phi$. This class of models can be thought as equivalent to perfect fluid models with zero vorticity and the pressure (Lagrangian) can be expressed through the energy density only~\cite{babi5}. 

Therefore, the three independent components are expressed as
\ben
\sigma &=&\frac{\G^{2}(\E)}{ r^{2}}\M_{v}=\frac{\G^{2}(\E)}{ r^{2}}\Big(m_{v}+2r\F^{2}(\E)\phi_{v}\phi_{vv}\Big), \label{47} \\
\r &=&\frac{\G^{2}(\E)}{ r^{2}}\M_{r}= \frac{\G^{2}(\E)}{ r^{2}}\Big(m_{r}+\F^{2}(\E)\phi_{v}^{2}\Big),\label{48} \\
\S &=& -\frac{\G^{2}(\E)}{ 2r}\M_{rr}=-\frac{\G^{2}(\E)}{ 2r}m_{rr}. \label{49}
\een

The energy conditions for the combination of Type-I and Type-II matter fields energy momentum tensor ${\cal T}_{\mu\nu}$, which are defined~\cite{haw-ellis,gm5} as following:

\textbf{(a)} The weak and strong energy conditions:
\be
\sigma\geq0~,~\rho\geq0~,~p\geq0~~~(\sigma\neq0).\label{50}
\ee

\textbf{(b)} The dominant energy conditions:
\be
\sigma\geq0~,~\rho\geq p\geq0  ~~~~~(\sigma\neq0).\label{51}
\ee

So, following the above energy conditions (\ref{50}) and (\ref{51}), imposed on ${\cal T}_{\mu \nu}$ will be constrained in $m(v,r)$ and $\phi(v)$ and their derivatives. Thus
\bea
\sigma > 0 \Rightarrow  m_{v}+2r\phi_v \phi_{vv} > 0\Rightarrow \frac{r^{1-2\kappa}}{1-2\kappa}\dot{f_{1}(v)}+\dot{f_{2}(v)}+2r\F^{2}(\E)\phi_{v}\phi_{vv}>0, \label{52} \\
\r > 0 \Rightarrow  m_{r}+\phi_{v}^{2}>0 \Rightarrow f_{1}(v)r^{-2\k}-\frac{1}{\G(\E)}\m^{2}c_{1}cr-\m^{2}c_{2}c^{2}+\F^{2}(\E)\phi_{v}^{2}>0~~\label{53} \\
p>0  \Rightarrow  m_{rr} < 0  \Rightarrow 2\kappa f_{1}(v)r^{-(1+2\kappa)}+\m^{2}c_{1}c>0.~~~\label{54}
\eea

\section{Collapsing scenario for the k-essence Vaidya spacetime in massive gravity's rainbow}
Now, let's cultivate the collapsing scenario of the {\bf k}-essence emergent Vaidya spacetime in the context of the massive gravity's rainbow. At the beginning, we define $K^{\mu}$ as the tangent to the non-spacelike geodesics with $K^{\mu}=\frac{dx^{\mu}}{d\l}$ where $\l$ is the affine parameter. The geodesic equation takes the form of~\cite{maombi,gm5}
\be
\bar{G}_{\mu\nu}K^{\mu}K^{\nu}=\b,\label{55}
\ee
where $\b$ is a constant. Here $\b=0$ and $\b<0$ describes the null geodesics and timelike geodesics respectively.

Expressing the {\bf k}-essence emergent geodesic equation (\ref{11}) in terms of $K^{\mu}$ we get
\be
\frac{dK^{\alpha}}{d\lambda}+\bar\Gamma ^{\alpha}_{\mu\nu}K^{\mu}K^{\nu}=0.\label{56}
\ee

Now, using Eqs. (\ref{35}), (\ref{55}) and (\ref{56}), we get the geodesic equations~\cite{gm5,joshi3,maombi} with the forms:
\ben
&&\frac{dK^{v}}{d\l}+\frac{\G(\E)}{2\F(\E)}\Big[\frac{\M}{r^{2}}-\frac{\M_{r}}{r}\Big](K^{v})^{2}- r\frac{\F(\E)}{\G(\E)}\Big[(K^{\t})^{2}+\sin^{2}\t(K^{\P})^{2}\Big]=0\nonumber\\
&\Rightarrow& \frac{dK^{v}}{d\l}+\frac{\G(\E)}{2\F(\E)}\Big[\frac{m}{r^{2}}-\frac{m_{r}}{r}\Big](K^{v})^{2}-\frac{\F(\E)}{\G(\E)}\frac{l^{2}}{r^{3}}=0,~\label{57}
\een

\ben
&&\frac{dK^{r}}{d\l}+\frac{\G(\E)}{2\F(\E)}\frac{\M_{v}}{r}(K^{v})^{2}-\frac{\G^{2}(\E)}{2}\Big[\frac{\M}{r^{2}}-\frac{\M_{r}}{r}\Big]\times \Big[-\frac{1}{\F^{2}(\E)}\Big(1-\frac{\M}{r}\Big)(K^{v})^{2}+\frac{K^{r}K^{v}}{\F(\E)\G(\E)}\Big]\nonumber\\&-&\Big(1-\frac{\M}{r}\Big)r\Big[(K^{\t})^{2}+\sin^{2}\t(K^{\P})^{2}\Big]=0\nonumber\\
&\Rightarrow& \frac{dK^{r}}{d\l}+\frac{\G(\E)}{2\F(\E)}\Big(\frac{m_{v}}{r}+2\F^{2}(\E)\phi_{v}\phi_{vv}\Big)(K^{v})^{2}-\frac{l^{2}}{r^{3}}\Big(1-\frac{m}{r}-\F^{2}(\E)\phi_{v}^{2}\Big)\nonumber\\&-&\frac{1}{2}\b\G^{2}(\E)\Big(\frac{m}{r^{2}}-\frac{m_{r}}{r}\Big)=0~,\label{58}
\een

\ben
&&\frac{dK^{\t}}{d\l}+\frac{2}{r}K^{\t}K^{r}-\sin\t~\cos\t(K^{\P})^{2}=0,
\label{59}
\een

\ben
&&\frac{d}{d\l}(r^{2}\sin^{2}\t K^{\P})=0,
\label{60}
\een
where we use the relation~\cite{joshi3,gm5}
\ben
K^{\P}=\frac{l\cos\d}{r^{2}\sin^{2}\t}~~,~~K^{\t}=\frac{l}{r^{2}}\sin\d \cos\P, \label{61}
\een
where $l$ and $\d$ are constant of integration. which represents the impact parameter and the isotropy parameter respectively, satisfying the relation, $\sin\P~\tan\d=\cot\t$.

The definition of $K^{v}$~\cite{joshi3,maombi} gives us
\ben
K^{v}=\frac{P(v,r)}{r}
\label{62}
\een
and the relation $\bar{G}_{\mu\nu}K^{\mu}K^{\nu}=\b$ provides
\ben
K^{r}&=&\frac{P}{2r}\frac{\G(\E)}{\F(\E)}\Big(1-\frac{m}{r}-\F^{2}(\E)\phi_{v}^{2}\Big)-\frac{\F(\E)}{\G(\E)}\frac{l^{2}}{2Pr}+\F(\E)\G(\E)\frac{\b r}{2P},
\label{63}
\een
where $P(v,r)$ is an arbitrary function.

Next, on differentiation the equation (\ref{62}) with respect to $\l$ we get
\ben
\frac{dP}{d\l} &=& \frac{1}{r}\Big(r^{2}\frac{dK^{v}}{d\l}+P\frac{dr}{d\l}\Big)\nonumber\\
&=& \frac{\G(\E)}{\F(\E)}\frac{P^{2}}{2r^{2}}\Big(1-\frac{2m}{r}+m_{r}\Big)+\frac{\F(\E)}{\G(\E)}\frac{l^{2}}{2r^{2}}+\F(\E)\G(\E)\frac{\b}{2}-\frac{\G(\E)}{\F(\E)}\frac{P^{2}}{2r^{2}}\F^{2}(\E)\phi_{v}^{2}. \label{64}
\een

After this we would now examine the destiny of the collapse, whether it ends with a black hole or a naked singularity for the given {\bf k}-essence emergent Vaidya massive gravity rainbow mass function (\ref{36}). If there exist some families of future directed non-spacelike trajectories which terminates in the past at the singularity, which are reaching faraway observers in spacetime, then a naked singularity forming as the final state of the collapse. On the flip side, if no such families exist and an event horizon forms sufficiently early to cover the singularity then we have a black hole.

The radial null geodesic $(l=0,~\beta=0)$ can be achieved using equations (\ref{62}) and (\ref{63}) as
\bea
\frac{dv}{dr}&&=\frac{2r\F(\E)}{\G(\E)\Big(r-m(v,r)-r\F^{2}(\E)\phi_{v}^{2}\Big)}\equiv\frac{2\F(\E)}{\G(\E)\Big(1-\frac{\M(v,r)}{r}\Big)}.
\label{65}
\eea

For a suitable choice of rainbow functions it has a singularity at $r=0,~v=0$, provided $\phi_{v}^{2}\neq 0$.

\subsection{Structure of the Central singularity for the {\bf k}-essence emergent Vaidya spacetime in massive gravity rainbow}
In general, the above Eq. (\ref{65}) can be written as~\cite{maombi,gm5}
\be
\frac{dv}{dr}=\frac{M(v,r)}{N(v,r)},
\label{66}
\ee
with the singular points $r=0,~v=0$, where both the above functions $M(v,r)$ and $N(v,r)$ vanish. 

As discussed in~\cite{maombi}, we can also have the characteristic equation for the existence and uniqueness of the above form of differential equation (\ref{66}) in the vicinity of the singularity as
\be
\chi^{2}-(A+D)\chi +AD-BC=0, \label{67}
\ee
where $A=M_{v}(0,0)~,~B=M_{r}(0,0)~,~C=N_{v}(0,0)~,~D=N_{r}(0,0)$ and $AD-BC\neq 0$. 

The roots of the above Eq. (\ref{67}) are
\be
\chi=\frac{1}{2}\Big[(A+D)\pm \sqrt{(A-D)^{2}+4BC}\Big].
\label{68}
\ee

The singularity can be of two type, First one is {\it node}, if $(A-D)^{2}+4BC\geq 0$ and $BC>0$. and second one is a center or focus.

Comparing Eqs. (\ref{65}) and (\ref{66}), we have
\bea
M(v,r) &=& 2r\F(\E),\nonumber\\ 
N(v,r) &=& \G(\E)\Big(r-\M(v,r)\Big) = \G(\E)\Big(r-m(v,r)-r\F^{2}(\E)\phi_{v}^{2}\Big).
\label{69}
\eea

At the central singularity ($v=0, r=0$), we define
\bea
\M_{0} = \lim\limits_{v\to 0, r \to 0}\M(v,r)~;~
\M_{v0} = \lim\limits_{v\to 0, r \to 0}\frac{\p}{\p v}\M(v,r)~;~\nonumber\\
\M_{r0} = \lim\limits_{v\to 0, r \to 0}\frac{\p}{\p r}\M(v,r)~;~
\phi_{v0} = \lim\limits_{v\to 0, r \to 0}\frac{\p}{\p v}\phi(v)~;~\nonumber\\
m_{v0} = \lim\limits_{v\to 0, r \to 0}\frac{\p}{\p v}m(v,r)~;~
m_{r0} = \lim\limits_{v\to 0, r \to 0}\frac{\p}{\p r}m(v,r).\label{70}
\eea

Using Eqs. (\ref{69}) and (\ref{70}), we get $A=0,~B=2\F(\E),~C=-\G(\E)m_{v0}~ and ~ D=\G(\E)\Big(1-m_{r0}-\F^{2}(\E)\phi_{v0}^{2}\Big)$. Hence, roots of the characteristic equation (\ref{67}) can be written as:
\bea
\chi=\frac{1}{2}\Bigg[\G(\E)\Big(1-m_{r0}-\F^{2}(\E)\phi_{v0}^{2}\Big) \pm \sqrt{\G^{2}(\E)\Big(1-m_{r0}-\F^{2}(\E)\phi_{v0}^{2}\Big)^{2}-8\F(\E)\G(\E)m_{v0}}\Bigg]. \label{71}
\eea

The required conditions for the singular point at $r=0,~ v=0$ to be a {\it node}
\bea
\G(\E)\Big(1-m_{r0}-\F^{2}(\E)\phi_{v0}^{2}\Big)^{2}\geq8\F(\E)m_{v0},~
m_{v0}>0~and~\phi_{v0}^{2}>0. \label{72}
\eea

Therefore, to satisfy the condition (\ref{72}) the {\bf k}-essence emergent Vaidya massive gravity rainbow mass function $\M(v,r)$ and the rainbow functions ($\F(\E)~and~\G(\E)$)can be chosen in a certain way and then the singularity at the origin $(v=0, r=0)$ will be a node resulting the outgoing non-spacelike geodesics come out of the singularity with a definite value of the tangent.

The null geodesic equation can be linearized near the central singularity using the limits (\ref{70}) as
\be
\frac{dv}{dr}=\frac{2r\F(\E)}{\G(\E)\Big[(1-m_{r0}-\F^{2}(\E)\phi_{v0}^{2})r-m_{v0}v\Big]}, \label{73}
\ee
which is also singular at $v=0,~r=0$ provided $\phi_{v}^{2}\neq 0$. From (\ref{36}), we get
\be
m_{r0}=-\m^{2}c_{2}c^{2};~m_{v0}=\frac{\p f_{2}(v)}{\p v}\Big{|}_{0}\equiv \dot{f_{20}(v)}.
\label{74}
\ee

\subsection{Existence of outgoing non-spacelike geodesics for the {\bf k}-essence emergent Vaidya spacetime in massive gravity rainbow}
The summary upto now is that, the outgoing radial null geodesics have to end up in the past at a singularity, particularly, in the central physical singularity located at $v=0,~ r = 0$. With the help of this remark, we will now explore the existence of naked singularity (NS) in the {\bf k}-essence emergent Vaidya spacetime in massive gravity rainbow. For a locally naked singularity, such null geodesics exist.  The possibility of this singularity to be either a naked singularity or a black hole (BH) implies that if the singularity is not a naked singularity, then the formation  of a black hole is undoubtable. So, the analysis of the radial null geodesics emerging from the singularity can help us to understand the nature of such a singularity, which can be established by a catastrophic gravitational collapse. 

It is well-known in general relativity that, such a singularity which comes out from a gravitational collapse, is always a black hole because of the cosmic censorship in general relativity. Therefore, there will be an event horizon by which the singularity is always covered in general relativity. But to speak about the generalized case, it is possible for inhomogeneous dust cloud to form a naked singularity through a collapse~\cite{eardley}. Also it is noted that some interesting results have been obtained for fluids whose equations of state (EOS) is not exactly similar to a dust cloud~\cite{joshi4}. Thus it becomes evident to generalize the cosmic censorship in general relativity~\cite{joshi5}.

Let us now consider that, the {\bf k}-essence emergent Vaidya massive gravity mass function $\M(v,r)$ with acceptable choices of rainbow functions satisfies all the physical energy conditions (\ref{50}) and (\ref{51}) with the constraints (\ref{52}), (\ref{53}) and (\ref{54}). The partial derivatives of the mass function which are continuous in the entire {\bf k}-essence emergent Vaidya massive gravity's rainbow spacetime (\ref{36}) also exist obeying the conditions (\ref{72}) at the central singularity. Following refs.~\cite{maombi,patil,heyd1,heyd2,gm5}, we would now determine the nature (a black hole or a naked singularity) of the collapsing solutions. To do this, we consider the function $X$ behave as, $X =\frac{v}{r}$ having limiting value at the central singularity as follows:
\be
X_{0}=\lim\limits_{v\to 0, r \to 0}X=\lim\limits_{v\to 0, r \to 0}\frac{v}{r}. \label{75}
\ee

Using Eqs. (\ref{65}),  (\ref{75}) and L'Hospital's rule, we get
\bea
X_{0}&&=\lim\limits_{v\to 0, r \to 0}\frac{v}{r}=\lim\limits_{v\to 0, r \to 0}\frac{dv}{dr}=\lim\limits_{v\to 0, r \to 0}\frac{2\F(\E)}{\G(\E)\Big(1-\frac{\M(v,r)}{r}\Big)}.  \label{76}
\eea

Again using (\ref{36}), the above Eq. (\ref{76}) can be written as
\bea
\frac{2}{X_{0}}=\lim\limits_{v\to 0, r \to 0}\frac{\G(\E)}{\F(\E)}\Big[1-\frac{r^{-2\k}}{1-2\k}f_{1}(v)-\frac{f_{2}(v)}{r} + \frac{1}{2\G(\E)}\m^{2}cc_{1}r+\m^{2}c^{2}c_{2} \nonumber\\-\F^{2}(\E)\phi_{v}^{2}\Big],\label{77}
\eea
under the constraint $\k \neq \frac{1}{2}$.
 
Now, considering $f_{1}(v)=\a v^{2\k}$ and $f_{2}(v)=\b v$, the algebraic equation of $X_{0}$ can be expressed as
\bea
\frac{\a}{1-2\k}X_{0}^{1+2\k}+\b X_{0}^{2}-\Big(1+\m^{2}c^{2}c_{2}-\F^{2}(\E)\phi_{v0}^{2}\Big)X_{0}+2\frac{\F(\E)}{\G(\E)}=0,~~~~~~~~\label{78}
\eea
 where $\a$ and $\b$ are constants. 

As mentioned before, the Eq. (\ref{78}) is not similar to  Eq. (3.7) of Heydarzade et al.~\cite{heyd2} due to the presence of the terms $\phi_{v0}^{2}$ and rainbow functions. But we choose the same functions of $f_{1}(v)$ and $f_{2}(v)$ as \cite{heyd2} along with the same characteristics. In the early universe ($\k \geq0$), $f_{1}(v)$ grows with $v$, whereas in the late universe ($\k < 0$), $f_{1}(v)$ decays with time. On the other hand $f_{2}(v)$ is a linear function of $v$. These choices of the arbitrary functions $f_{1}(v)$ and $f_{2}(v)$ are self-similar in nature and this comes from the definition of $X_{0}$ given in Eq. (\ref{76}). For the same reasons, showed by Heydarzade et al.~\cite{heyd2}, we also consider self-similar cases.

The formation of a black hole can be assured in this system by achieving a non-positive solution of the above Eq. (\ref{78}). However, the positive roots of this equation can produce {\it a naked singularity}. With the specific choice of the rainbow functions, we will examine the effects of the mass term and kinetic energy of the {\bf k-}essence scalar field ($\phi_{v0}^{2}$) on the formation of naked singularity and observe that the effect of a mass term and $\phi_{v0}^{2}$ on the formation of naked singularities. For this purpose, we choose a specific rainbow function~\cite{camel4,camel3,heyd1} as
\be
\F(\E)=1,~\G(\E)=\sqrt{1-\eta \Big(\frac{E_{1}}{E_{p}}\Big)}. \label{79}
\ee

In the above expressions, we have denoted Planck energy by $E_{p}$, given by $E_{p}=1/\sqrt{G}=1.221\times 10^{19}$GeV, where G denotes gravitational constant and $E_{1}=1.42\times 10^{-13}$~\cite{camel4,camel3,heyd1}. In \cite{camel3,heyd1}, the authors have estimated the value of $\eta$ as, $\eta\approx 1$. So, in our study we use $\eta=1$.  It is very cumbersome to find exact solutions for $X_{0}$, except for some particular values. Therefore, for some preferable values we find the exact solutions as:\\

{\it Case-I: $\k=0$:} The zero value of $\k$ represents the pressure-less dust regime of the universe. For this regime, the Eq. (\ref{78}) reduces to
\be
\a X_{0}+\beta X_{0}^{2}-\Big(1+\m^{2}c^{2}c_{2}-\F^{2}(\E)\phi_{v0}^{2}\Big)X_{0}+2\frac{\F(\E)}{\G(\E)}=0,\label{80}
\ee
and the solutions become
\ben
X_{0\pm}&&=\frac{\Big(U-\F^{2}(\E)\phi_{v0}^{2}-\a\Big)}{2\b}  \pm\frac{\sqrt{(U-\F^{2}(\E)\phi_{v0}^{2}-\alpha)^{2}-8\b\F(\E)/\G(\E)}}{2\b}~~~
\label{81}
\een
with $U=1+\m^{2}c_{2}c^{2}$.

For the demand of these two solutions to be real and finite, we must have $\b\neq 0$ and $\G(\E)(U+\F^{2}(\E)\phi_{v0}^{2}-\alpha)^{2}\geq 8\beta\F(\E)$. In our case the values of $\phi_{v0}^{2}$ are $0<\phi_{v0}^{2}<1$, i.e., positive and the rainbow functions (\ref{79}) are also positive.

To make the solution $X_{0+}$ positive, we have the conditions for:\\ 

(i) $\beta>0:~~ (U-\F^{2}(\E)\phi_{v0}^{2}-\alpha)\geq
-\sqrt{(U-\F^{2}(\E)\phi_{v0}^{2}-\a)^{2}-8\beta\F(\E)/\G(\E)}\\~~\Rightarrow~\beta\geq 0$. 

But if the solutions are real and finite $\b\neq 0$, then the solution $X_{0+}$ is positive for any positive $\b$.\\

(ii) $\beta<0:~~~(U-\F^{2}(\E)\phi_{v0}^{2}-\alpha)\leq -\sqrt{(U-\F^{2}(\E)\phi_{v0}^{2}-\alpha)^{2}-8\beta\F(\E)/\G(\E)}\\~~\Rightarrow ~~\beta\leq 0$.\\

As $\beta\neq 0$, the solution $X_{0+}$ has positive values for any negative $\b$. Therefore, in conclusion we can say that, for any non-zero real values of $\beta$, the solution $X_{0+}$ is positive and this represents a naked singularity.

For positivity of the solution $X_{0-}$, we have, for $\beta>0$, \\ $(U-\F^{2}(\E)\phi_{v0}^{2}-\alpha)\geq \sqrt{(U-\F^{2}(\E)\phi_{v0}^{2}-\alpha)^{2}-8\beta\F(\E)/\G(\E)}~\Rightarrow~\b\geq 0$. \\ For the above said reason as explained above, here also $\b>0$, which implies that the solution is positive. Again for $\beta<0$, we can get the positive solution following the same reason. Hence, we can conclude that, this solution also represents a naked singularity. Finally, for $\kappa=0$, we will get naked singularity as the destiny of the collapse. Here, we would like to mention that, the conditions for the positive solutions of Eq. (\ref{78}) are similar type as Heydarzade et al.~\cite{heyd2}.\\

{\it Case-II: $\kappa=1$:} This condition represents the early stiff fluid~\cite{Zeldovich} era of our universe. In this scenario Eq. (\ref{78}) reduces to
\be
-\alpha X_{0}^{3}+\beta X_{0}^{2}-(U-\F^{2}(\E)\phi_{v0}^{2})X_{0} +2\frac{\F(\E)}{\G(\E)}=0. \label{82}
\ee

If we consider $\alpha$, $\beta$, $U-\F^{2}(\E)\phi_{v0}^{2}$, $\F(\E)$ and $\G(\E)$, all to be positive, then by the well-known theorem of Descartes, also known as Descartes' Rule of Sign, the number of variations of sign is 3. Which means the number of positive roots of Eq. (\ref{82}) is either 1 or 3. The minimum number of  positive root being 1, we can announce that the naked singularity forms for $\kappa=1$.\\

It is noteworthy, for the case $\kappa>\frac{1}{2}$, the energy conditions (50-52) are violated apparently in our case and these energy conditions can be made satisfied easily if we redefine the mass function $\M(v, r)$.

{\it Case-III: $\kappa=-1/2$:} This represents the dark energy dominated accelerating universe. For $\k=-1/2$, Eq.  (\ref{78}) is reduces to
\be
\frac{\alpha}{2}+\beta X_{0}^{2}-(1+\mathbb{M}^{2}c_{2}c^{2}-\F^{2}(\E)\phi_{v0}^{2})X_{0} +2\frac{\F(\E)}{\G(\E)}=0, \label{83}
\ee
which bears the following solutions
\bea
&&X_{0\pm}=\frac{1}{2\beta}\Big[(U-\F^{2}(\E)\phi_{v0}^{2})\pm\sqrt{(U-\F^{2}(\E)\phi_{v0}^{2})^{2}-\frac{2\beta}{\G(\E)}\Big(4\F(\E)+\alpha \G(\E)\Big)}\Big].~~~~\label{84}
\eea

In this case, to get the real and finite solutions, we must have $\beta \neq 0$ and $(U-\F^{2}(\E)\phi_{v0}^{2})^{2}\geq \frac{2\beta}{\G(\E)}(4\F(\E)+\alpha \G(\E))$. Following the same procedure of case-I, the positiveness of these solutions (\ref{84}), we can show for $\beta>0$, we get the condition $\alpha \frac{\G(\E)}{\F(\E)}\geq -4$ and for $\beta<0$, $\alpha \frac{\G(\E)}{\F(\E)}\leq -4$. Due to the presence of the rainbow functions, the solutions we get here are different from Heydarzade et al.~\cite{heyd2}.

\subsection{Numerical solution of $X_{0}$}
Using numerical methods, we will try to find a numerical solutions of $X_{0}$ by considering the specific rainbow functions (\ref{79}) in this subsection. The dynamics of collapse lies in the knowledge of $X_{0}$ for the whole cosmologically meaningful region $\kappa < 1$, i.e., from early to late universe, not only at discrete points of $\k$. So, we decide to visualize these solutions obtaining various contours for $\k-X_{0}$ for different numerical values of the involved parameters. 

\begin{figure}[h]
\centering
\includegraphics[width=8.0cm]{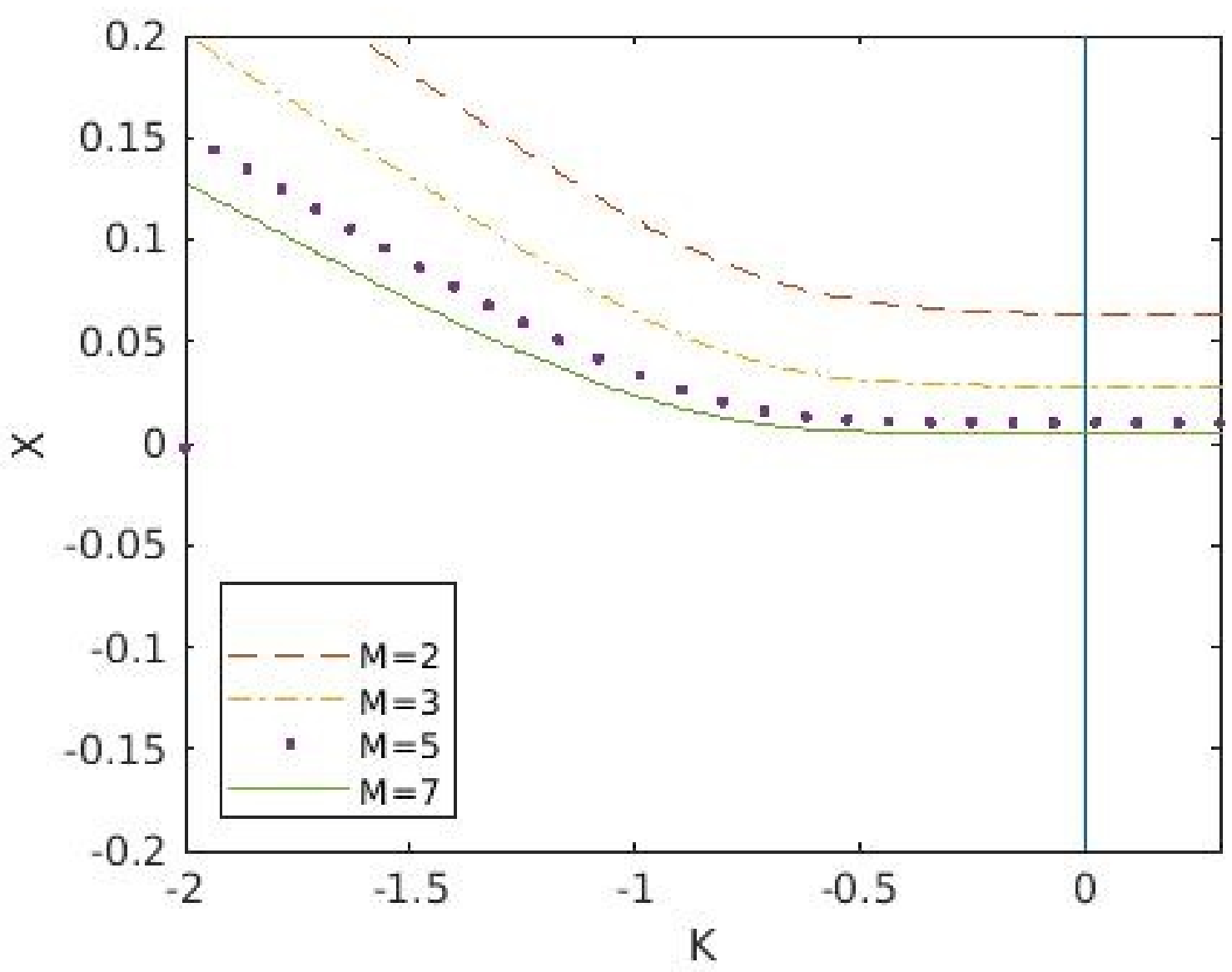}
\includegraphics[width=8.0cm]{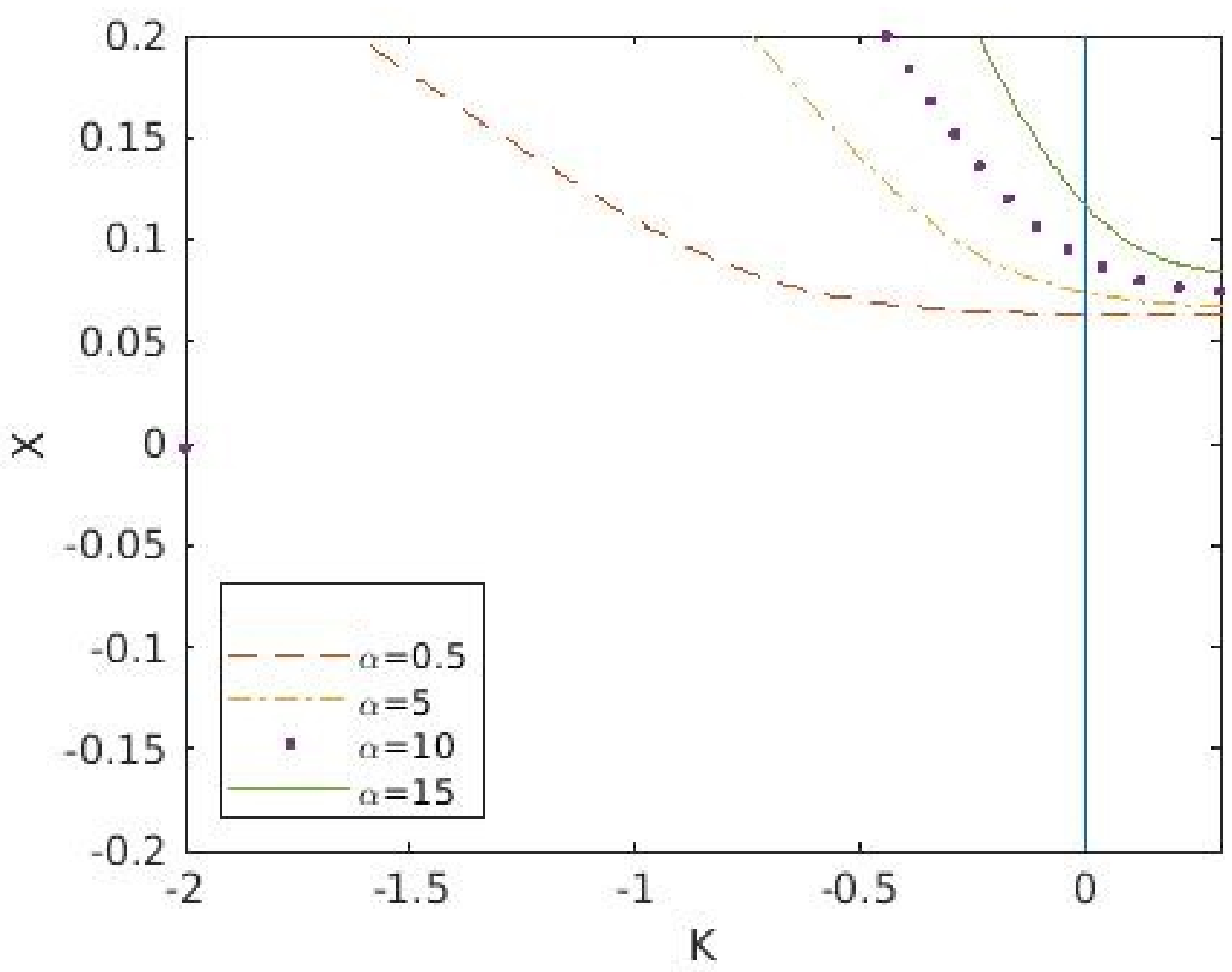}
\caption{Figs. (a) and (b) show the variation of $X_{0}$ with $\kappa$ for different values of $\mathbb{M}$ and $\alpha$ respectively (in Fig. (a) $M$ are equal to $\mathbb{M}$) with $\phi_{v0}^{2}=0.7$, $E_{p}=1.221\times 10^{19}$ and $E_{1}=1.42\times 10^{-13}$. In Fig. (a) the initial conditions are $\alpha=0.5$, $\beta=2$, $c=2$ and $c_{2}=2$ whereas in Fig. (b) the initial conditions are $\mathbb{M}=2$, $\beta=2$, $c=2$ and $c_{2}=2$.}
\label{fig:test}
\end{figure}

\begin{figure}[h]
\centering
\includegraphics[width=8.0cm]{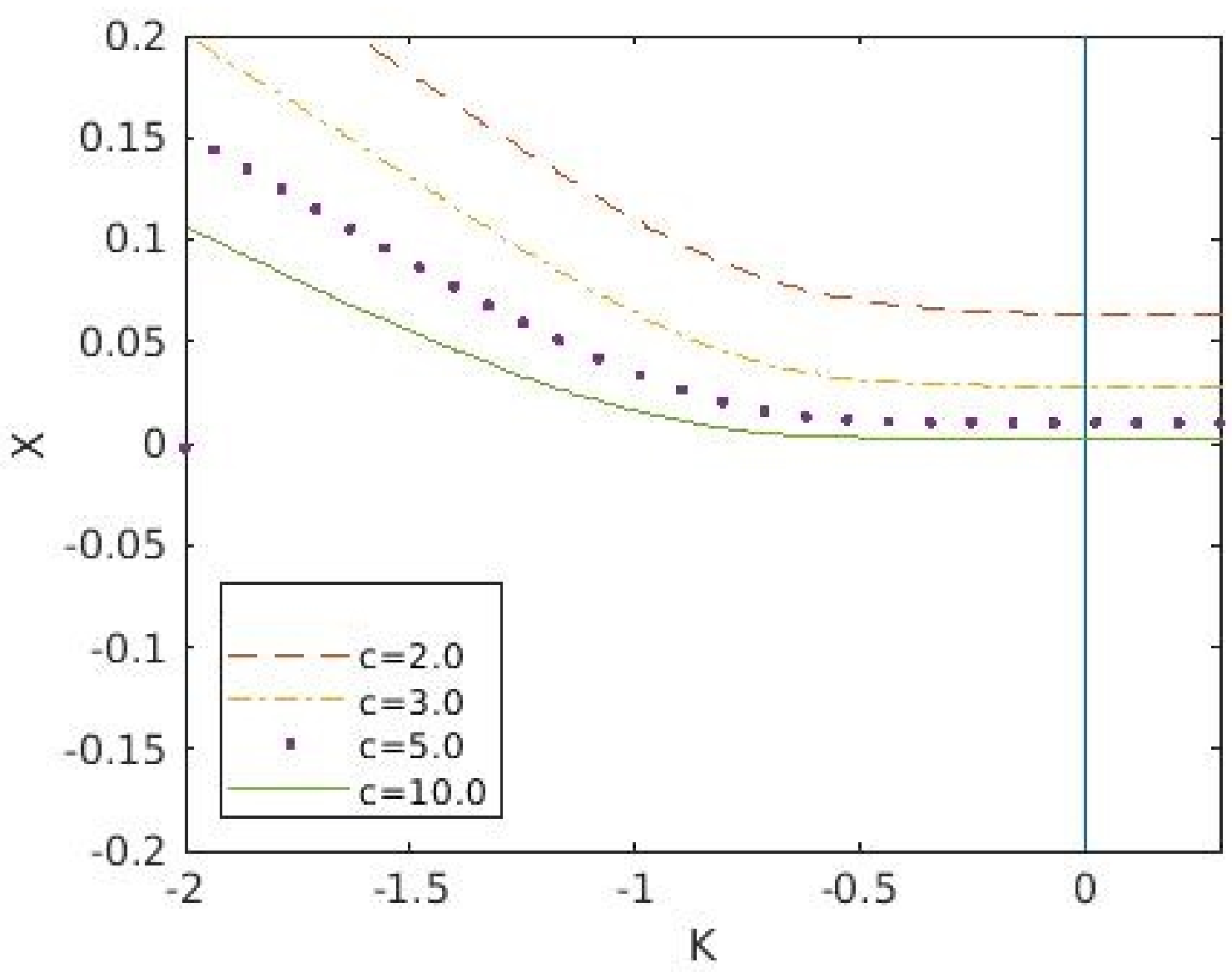}
\includegraphics[width=8.0cm]{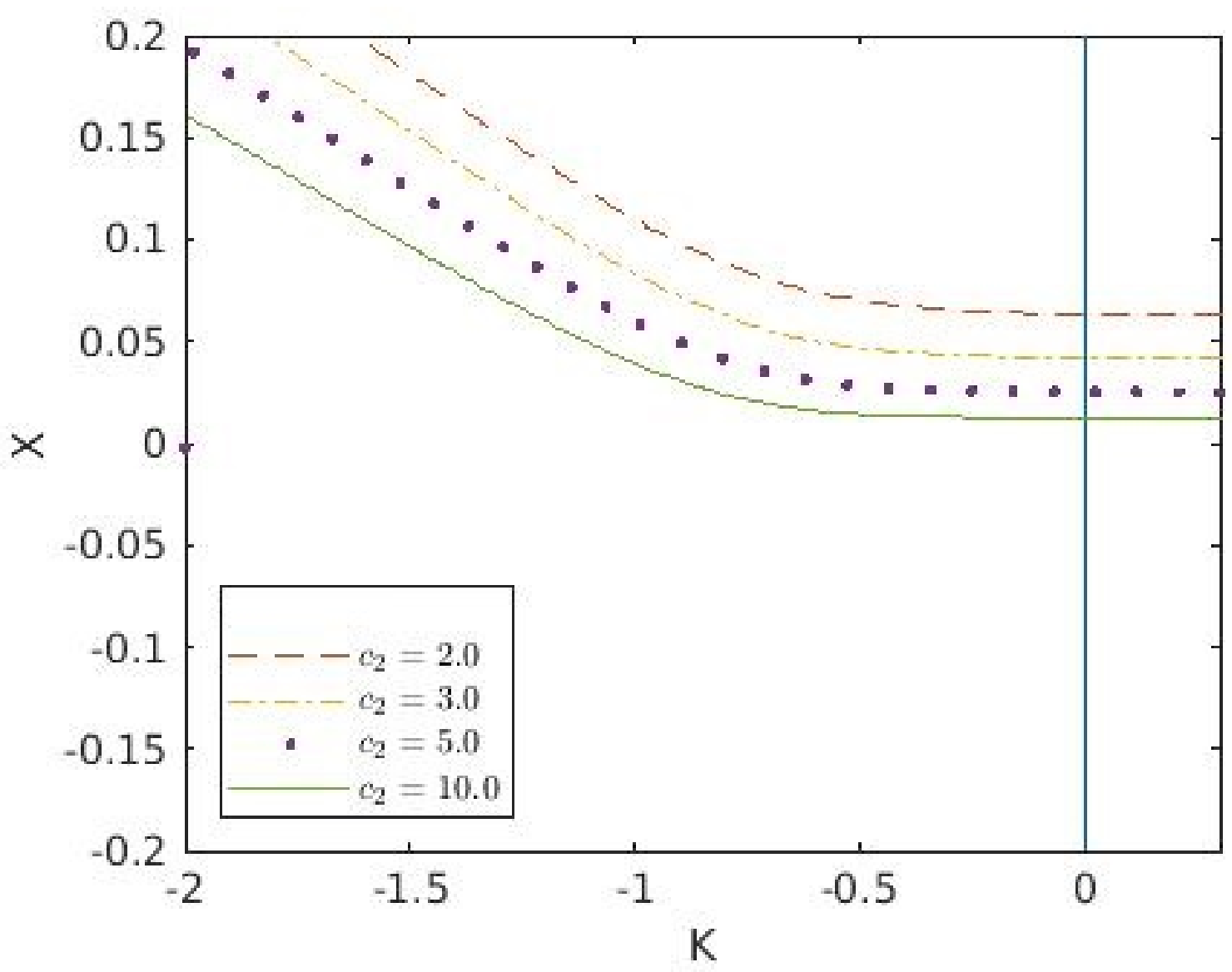}
\caption{Figs. (a) and (b) show the variation of $X_{0}$ with $\k$ for different values of $c$ and $c_{2}$ respectively with $\phi_{v0}^{2}=0.7$, $E_{p}=1.221\times 10^{19}$ and $E_{1}=1.42\times 10^{-13}$. In Fig. (a) the initial conditions are $\mathbb{M}=2$, $\alpha=0.5$, $\beta=2$ and $c_{2}=2$ whereas in Fig. (b) the initial conditions are $\mathbb{M}=2$, $\alpha=0.5$, $\beta=2$ and $c=2$.}
\label{fig:test}
\end{figure}

\begin{figure}[h]
 \centering
 \includegraphics[width=8.0cm]{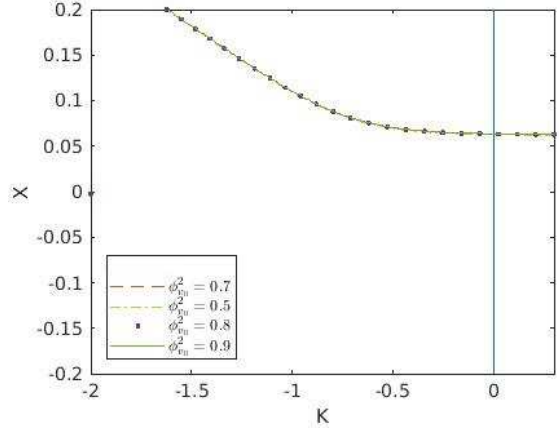}
  \caption{Variation of $X_{0}$ with $\kappa$ for different values of $\phi_{v0}^{2}$ with $E_{p}=1.221\times 10^{19}$ and $E_{1}=1.42\times 10^{-13}$ and the initial conditions are $\alpha=0.5$, $\beta=2.0$, $c=2.0$, $c2=2.0$ and $\mathbb{M}=2.0$}
\end{figure}

Looking at the above figures, we get to know that, the trajectories run across the positive range of $X_{0}$, which is the indication of formation of naked singularity (NS). Fig. 1(a) reveals the dependence of $X_{0}$ on the equation of state (\ref{23}) parameter ($\kappa$) for particular values of the massive gravity parameter $\m(=2,3,5,7)$ with the fixed $\phi_{v0}^{2}$. Also, from this figure, we see that with increasing $\mathbb{M}$ the tendency of formation of NS decreases. Therefore, we can say that the addition of graviton mass to this system with the specific choice of rainbow functions (\ref{79}) deforms the dynamics of the system. In Fig. 1(b), we obtain the $\kappa-X_{0}$ trajectories by varying the values of $\alpha(=0.5,5,10,15)$. In this figure, we see that, with greater values of $\a$, the tendency to form NS is greater.

In Fig. 2(a), we observe how $c$ affects the collapsing scenario. It is clear that, for the greater value of $c(=2,3,5,10)$, the tendency to form NS is lesser. In Fig. 2(b), we can observe the effect of $c_{2}$ on the collapsing system. Like $c$, here also we see that an increase in $c_{2}$ decreases the possibility of NS.

In Fig. 3, we have showed the effect of kinetic energy ($\phi_{v0}^{2}$) of the {\bf k}-essence scalar field on the collapsing scenario. It has been observed that, all the values of $\phi_{v0}^{2}$ predict the same trajectories of the system and form the NS of the collapsing system. Here we consider restriction of  $\phi_{v0}^{2}$ not to exceed unity.

From all the above observations and their corresponding discussions of the $\kappa-X_{0}$ trajectories makes it clear that the trajectories reside in the positive $X_{0}$ region, and assures the formation of naked singularity from the collapse of this system.

\subsection{Strength of singularity for the {\bf k}-essence emergent Vaidya spacetime in massive gravity rainbow}
Taking the recommendation of~\cite{gm5,heyd2,patil,tipler,ghosh,clarke}, we can conclude that, a singularity $(r=v=\l=0)$ would be strong if the following condition is satisfied
\be
\lim\limits_{\l\to 0}\l^{2}\psi=\lim\limits_{\l\to 0}\l^{2}\bar{R}_{\mu\nu}K^{\mu}K^{\nu}>0,
\label{85}
\ee
where $\bar{R}_{\mu\nu}$ is the Ricci tensor and $\psi=\bar{R}_{\mu\nu}K^{\mu}K^{\nu}$ is defined as a scalar of the {\bf k-}essence emergent Vaidya massive gravity rainbow spacetime. We would like to mention that, this scalar $\psi$ is not the {\bf k-}essence scalar field and is not coupled with the background gravitational metric $g_{\mu\nu}$. Stepping into the footsteps of \cite{gm5,maombi}, with the conditions of (\ref{72}), it can be shown that
\be
\lim\limits_{\l\to 0}\l^{2}\psi=\G^{2}(\E)\Big(\M_{v0}\Big)\frac{1}{4}X_{0}^{2}>0. \label{86}
\ee

So far we have seen that, if this condition is satisfied for some real and positive roots of $X_{0}$, the naked singularity of the {\bf k-}essence emergent Vaidya massive gravity rainbow spacetime is strong. On the other hand, if there is no positive real root of $X_{o}$, we can conclude that there is no outgoing future directed null geodesics from the singularity exist, i.e., the collapse ends in a black hole.

Now from the definition of $f_{1}(v)$ and $f_{2}(v)$ and from (\ref{36}), we have
\be
\mathcal{M}_{v}=\frac{2\k\a}{1-2\k}\Big(\frac{v}{r}\Big)^{2\k-1}+\b+2r\F^{2}(\E)\phi_{v}\phi_{vv}. \label{87}
\ee

Thus, we have
\be
\mathcal{M}_{v0}=\frac{2\k\a}{1-2\k}X_{0}^{2\k-1}+\b\equiv m_{v0}. \label{88}
\ee

In the above three cases, i.e., $\k=0$, $\k=1$ and $\k=-1/2$, we have achieved the positive roots of $X_{0}$ for the respective conditions. Also, from Eq. (\ref{88}), we observe the following situations for the above three cases:\\

(i) $\k=0$, $\mathcal{M}_{v0}=\b$ which is positive for any positive value of $\b$,

(ii) $\k=1$, $\mathcal{M}_{v0}=\a X_{0}+\b$, positivity of $\mathcal{M}_{v0}$ implies that $\a X_{0}+\b>0$ and

(iii) $\k=-1/2$, $\mathcal{M}_{v0}=\b-\frac{\a}{2X_{0}^{2}}$, here the positive value of $\mathcal{M}_{v0}$ implies that $2\b X_{0}^{2}>\a$.

With the above conditions and the specific rainbow functions (\ref{79}), we see that $\lim\limits_{\l\to 0}\l^{2}\psi>0$. Therefore, it can be admitted that, the naked singularity is strong with the above conditions.

\subsection{Apparent horizon for the emergent spacetime and physical consequences}

\subsubsection{Apparent horizon}
The authors in~\cite{gm4,gm6}, have described the dynamical horizon (DH) on the basis of~\cite{dh1,dh2,dh3,dh4,dh5} in the context of {\bf k-}essence geometry considering Schwarzschild metric as the background metric. Whereas, in~\cite{maombi,gm5}, the authors have described the apparent horizon (AH) for the generalized Vaidya type geometry. The AH is nothing but the boundary of the trapped surface region in the given spacetime (\ref{35}).The casual behaviour of the trapped surfaces developed in the spacetime during the collapse evolution decides the occurrence of a naked singularity or black hole.

We should remember that, not only the AHs are not invariant properties of a spacetime, but also are distinct from event horizons (EHs). Within an AH, light does not move away from it. This is in contrast with the EH, where in a dynamical spacetime outgoing light rays exterior to an AH (but still interior to the EH) can exist. Specifically, an AH is a local notion of the boundary of a spacetime, whereas an EH is a global notion of a black hole.

\subsubsection{Physical consequences}
The AH for the {\bf k-}essence emergent Vaidya spacetime in massive gravity rainbow (\ref{35}) can be represented as
\ben
\frac{\M(v,r)}{r}=1\Rightarrow \frac{m(v,r)}{r}+\F^{2}(\E)\phi_{v}^{2}=1, \label{89}
\een
with $\F(\E)\neq 0$.

Considering refs.~\cite{maombi,gm5}, the slope of the AH at the central singularity ($r\rightarrow 0, v\rightarrow 0$) can be written as
\ben
\big(\frac{dv}{dr}\big)_{AH}&&=\frac{1-m_{r0}-\F^{2}(\E)\phi_{v0}^{2}}{m_{v0}}=\frac{1+\m^{2}c_{2}c^{2}-\F^{2}(\E)\phi_{v0}^{2}}{\dot{f}_{20}(v)},\label{90}
\een
where we have used Eq. (\ref{74}). 

Now, the sufficient conditions for the existence of a locally naked singularity for the collapsing {\bf k-}essence emergent Vaidya spacetime in massive gravity's rainbow has been attained. The emergent mass function $\M(v,r)$ obeys all the energy conditions and constraints mentioned in (\ref{52}), (\ref{53}) and (\ref{54}). Also, like ~\cite{gm5}, there exist an open set of parameter values for which, the singularity is locally naked for the case of {\bf k-}essence emergent Vaidya spacetime in massive gravity's rainbow.

Not to forget about the {\bf k-}essence emergent Vaidya spacetime in massive gravity rainbow metric may exchange radiation with the surrounding but the mass function (\ref{36}) can not totally evaporate due the presence of massive gravity terms and the kinetic part of the {\bf k-}essence scalar field, which is function of $v$. As an example, in~\cite{gm6}, the authors have shown the decreasing nature of black hole mass $m(v,r)$ in the {\bf K-}essence emergent Schwarzschild Vaidya spacetime but it does not completely vanish, by using DH equation as $\phi_{v}^{2}\rightarrow 0^{+}$.

\section{Discussion and Conclusion}
This paper shows the construction of a theory of the {\bf k}-essence Vaidya spacetime in massive gravity’s rainbow. We have also analyzed the energy dependent deformation of the massive gravity in our work. We have used the Vainshtein mechanism and the dRGT mechanism in the construction of massive gravity where, deformation by the rainbow functions has been considered. This paper also includs the analysis of the gravitational collapse of null fluids in the radiating {\bf k}-essence Vaidya black hole solution of massive gravity’s rainbow based on the Dirac-Born-Infeld Lagrangian. The above said exploration has been presented in the context of cosmic censorship hypothesis with the {\bf k}-essence emergent Vaidya massive gravity rainbow mass function. The Einstein tensor and the energy conditions for the combination of Type-I and Type-II matter fields energy-momentum tensor for the {\bf k}-essence emergent Vaidya massive gravity rainbow spacetime has also been studied thoroughly. We have shed some light on the classification of non-spacelike geodesic for the {\bf k}-essence emergent Vaidya massive gravity rainbow spacetime, that connects the naked singularity in the past, which is a strong curvature singularity in a stronger sense. The central singularity is a node, which allows outgoing non-spacelike geodesics to come out of the singularity pursuing a definite value of the tangent. This positivity of the tangent implies the violation of the strong cosmic censorship conjecture, i.e., the singularity is locally naked, though not necessarily the weak one.

If somehow, the value of tangent does not seizes positive value, the central singularity deviates from being a naked one because, in that case, there are no outgoing future-directed null geodesics coming out from the singularity, i.e., the collapse will always lead to a black hole. With these results we have also studied the effects of the graviton mass, kinetic energy of the k-essence scalar field and rainbow deformation for a time-dependent system.

We have obtained the contours for the tangent $X_{0}$ against the barometric parameter $\k$ for different values of other parameters, namely $\m$, $\a$, $\b$, $\phi_{v0}^{2}$, etc. Various regimes of the fluid content of the universe such as the radiation ($\k > 0$), pressure-less dust ($\k = 0$), dark energy ($\k < 0$) has been plotted. From those figures it is clear that the negative region contains no trajectories for specific choice of the rainbow functions. This outcomes make us believe the existence of black holes cannot be possible as the end state of collapse in the context of massive gravity rainbow with the considered parameters.

It should be noted that here we have studied the structure of gravitational collapse of a non-rotating type spacetime. The violation of CCC allows us to achieve a naked singularity which by definition is not bounded by any EH. On the other hand, there have been considered several models when the non-spacelike geodesics originate in the central singularity. However, this is not the prove of the nakedness of the singularity. In \cite{Grib2}, for the rotating case there should be a consideration of ergosphere to describe the motion of the non-spacelike geodesics.  The authors also have studied in detail the geodesics in the ergosphere of rotating black hole which have negative energies. An ergoshpere is basically a region located outside a rotating black hole's outer EH. Particularly in non-extremal Kerr spacetime ($a < M$ case), a non-spacelike geodesic for particles having negative, zero and positive energy can originate at the central singularity, when the collapse is a black hole. In \cite{Grib1} the authors have studied the dynamics of the high energy particles which have negative and positive energies in the vicinity of black holes. They also have studied the time of movement of particles with negative energy in the ergosphere. Vertogradov in his paper \cite{Verto} has examined the nature of the geodesics for particles with negative energy in Kerr’s metric. Also, we would like to mention that in our article we have not discussed about the formation of ergosphere. We may try to explore this type of formation in near future.

In the paper we have found very much similarity between the {\bf k}-essence emergent gravity metric and the generalized Vaidya massive gravity metric with the rainbow deformations. Obviously, this may establish another motivation or foundation for gravity's rainbow theory. The result continues to include that, the rainbow deformations of the {\bf k}-essence emergent Vaidya massive gravity metric also satisfies the required energy conditions. Therefore, from totally physical background, the results are viable and it demands for some observational clues on the possibility to observe the naked singularity. The crucial issue then becomes: How experimentally the naked singularity can be tested? Let us elaborate this point a bit based on the available literature as follows:

(i) The authors of~\cite{galin}, have studied the optical appearance of a geometrically thin and optically thick accretion disk around the strongly naked Janis-Newman-Winicour (JNW) singularity in their paper. Their solution represents an example of a compact object spacetime, which possesses no photon sphere. They~\cite{galin} also have established a general observational signature for distinguishable compact objects possessing no photon sphere from black holes. 

(ii) In~\cite{Shaikh,Shaikh1}, the authors have studied the distinction between black holes and naked singularities using the images of their accretion disks. They have considered a simplified model of spherical accretion onto the central object and studied shadow and images of Joshi-Malafarina-Narayan (JMN)~\cite{joshi6,joshi7} for naked singularities.

(iii) Bhattacharya et al.~\cite{Bhattacharya} have studied a new class of naked singularities and their observational signatures on the basis of suitable junction conditions matched with an expanding FLRW metric with a generic contracting solution of Einstein’s equations, on a spacelike hypersurface. They have also studied some observational aspects of the resulting two-parameter family of static naked singularity backgrounds. 

(iv) Kovacs et al.~\cite{Kovacs} have noted the properties of the accretion disk and observationally different black holes from naked singularities. They have considered a rotating solution of the Einstein-massless scalar field equations for the naked singularity, which reduces to the Kerr solution when the scalar charge tends to 1. They also have showed their interest to study the motion of the particles in the gravitational potential of this solution. Depending on the values of the mass, scalar charge, and the spin parameter, respectively, there are two types of disks that could exist around naked singularities. The first type are marginally stable orbits, locates outside the naked singularity, while the second type are reachable for the particles and in direct contact with the singularity. 

(v) Ortiz et al.~\cite{Ortiz} have spent their time to study the observational differences between black holes and naked singularities through the redshift function. They said that, the photons propagating from past to future null infinity through the center of the cloud  obtains the frequency shift, which can be measured by distant stationary observers and that can help to discriminate naked singularities from black holes. They have showed that, according to their model, it is possible to differentiate the formation of a naked singularity from the formation of a black hole based on the detection of photons traversing a collapsing object. 

(vi) Vertogradov~\cite{Vertogradov,Vertogradov1} have proposed a method to study the formation of the eternal naked singularity in the case of gravitational collapse of generalized Vaidya spacetime. Also, in~\cite{Virbhadra,Shaikh2}, the authors have studied the naked singularity, using the gravitational lensing. 

We would like to end our discussion by mentioning that, we have analytically discussed the formation of naked singularity in the {\bf k-}essence Vaidya spacetime in massive gravity’s rainbow. But from the above observational discussions of the existence of naked singularity, one may conclude that this type of singularity can also be possible to exist in reality, however which needs observational assurance and evidences.\\

{\bf Data availability:} Our manuscript has no associated data.

\section*{Acknowledgement}
The authors would like to thank the referee for illuminating suggestions to improve the paper.
SR is thankful to the Inter-University Centre for Astronomy and Astrophysics (IUCAA), Pune, India for providing Visiting Associateship under which a part of this work was carried out.\\

\end{document}